\pdfoutput=1

\documentclass[12pt]{iopart}

\usepackage[latin1]{inputenc}
\usepackage[dvips]{graphicx}
\usepackage{bm}
\usepackage{hyperref}
\usepackage{ulem}

\usepackage{color}
\definecolor{greencolor}{rgb}{0,0.5,0.2}
\definecolor{redcolor}{rgb}{.7,0.,0.}
\definecolor{bluecolor}{rgb}{0,0.,1.}
\definecolor{greycolor}{rgb}{.5,.5,.5}

\begin{document}

\title{Identification of Literary Movements Using Complex Networks to Represent Texts}

\author{Diego Raphael Amancio$^1$, Osvaldo N. Oliveira Jr.$^1$, Luciano da Fontoura Costa$^1$}

\address{$^1$  Institute of Physics of S\~ao Carlos \\
	University of S\~ao Paulo, P. O. Box 369, Postal Code 13560-970 \\
	S\~ao Carlos, S\~ao Paulo, Brazil \\
}


\ead{diego.amancio@usp.br,diegoraphael@gmail.com}

\begin{abstract}
    The use of statistical methods to analyze large databases of text has been useful to unveil patterns of human behavior and establish historical links between cultures and languages. In this study, we identify literary movements by treating books published from 1590 to 1922 as complex networks, whose metrics were analyzed with multivariate techniques to generate six clusters of books. The latter correspond to time periods coinciding with relevant literary movements over the last 5 centuries. The most important factor contributing to the distinction between different literary styles was {the average shortest path length (particularly, the asymmetry of the distribution)}. Furthermore, over time there has been a trend toward larger average shortest path lengths, which is correlated with increased syntactic complexity, and a more uniform use of the words reflected in a smaller power-law coefficient for the distribution of word frequency. Changes in literary style were also found to be driven by opposition to earlier writing styles, as revealed by the analysis performed with geometrical concepts. The approaches adopted here are generic and may be extended to analyze a number of features of languages and cultures.
\end{abstract}

\newpage
\tableofcontents

\maketitle

\section{Introduction}

Many findings related to language and culture issues have been made with the use of statistical methods to treat large amounts of texts~\cite{googleBook,twitter,metaknowledge,gb2}. Recent examples are the analysis of millions of books~\cite{googleBook} and the study of twitter messages, where the global variation of mood could be observed through textual analysis of tweets~\cite{twitter}. In several of such examples knowledge is inferred from the analysis of semantic contents in the texts. There are also other methods to analyze text, including cases where text is represented as a graph (or network)~\cite{newman}. Of particular relevance was the finding that networks formed from texts are scale free~\cite{beyond}, whose topology could be analyzed leading to various contributions. For instance, the scale-free structure (which is analogous to the Zipf's Law frequency distribution~\cite{corrr}) of text networks emerged as a consequence of an optimization process for both hearer and speaker, so that the effort to transmit and obtain a message was minimized~\cite{cancho2}. In addition to allowing for cultural features to be identified and explored, automatic analysis may be useful for real-world applications, such as automatic text summarization~\cite{summ}, machine translation~\cite{trad1,trad2}, authorship attribution~\cite{altmann}, information retrieval~\cite{infretr} and search engines~\cite{searEngine}.

In this study we used topological metrics of complex networks representing text from 77 books dating from 1590 to 1922 in an attempt to verify changes in writing style. With multivariate statistical analysis of the metrics obtained, we were able to identify periods that correspond to major literary movements. Furthermore, we established which network characteristics were responsible for the changes in writing style.

\section{Modeling Texts as Complex Networks}  \label{modelamento}

\subsection{Pre-Processing}

\begin{figure}
\begin{center}
\includegraphics[width=0.8\textwidth]{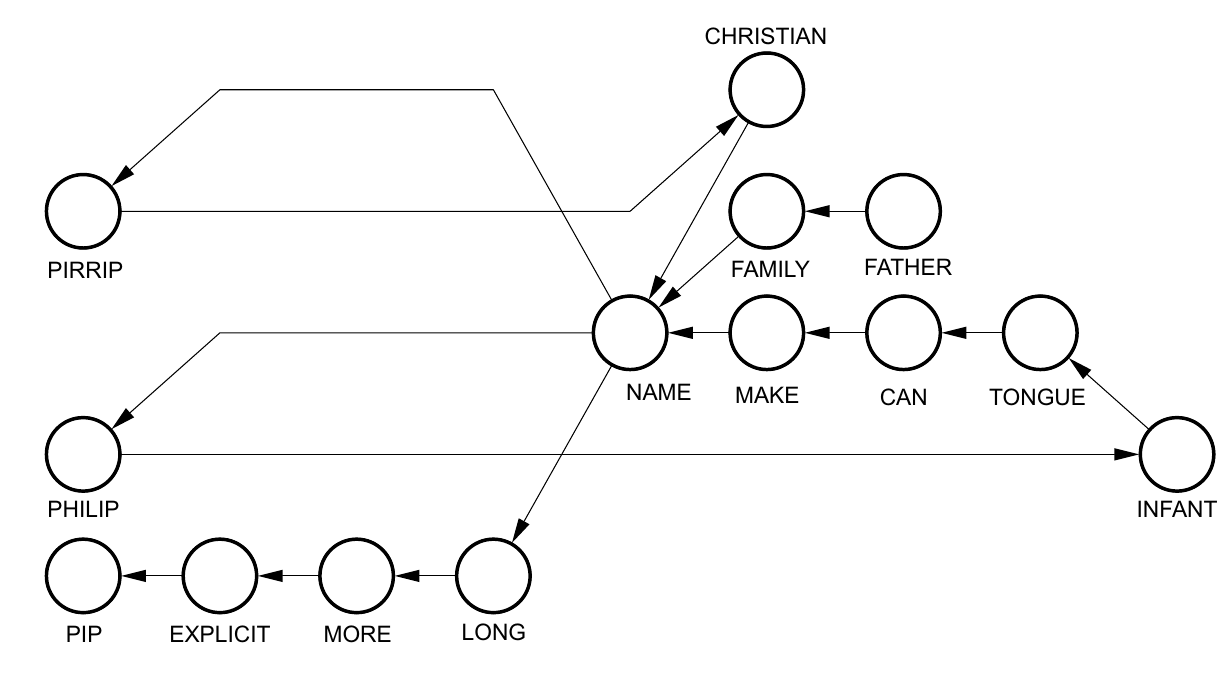}
\end{center}
\caption{\label{fig:network}Network obtained from the extract ``My father's family name being Pirrip, and my Christian name Philip, my infant tongue could make of both names nothing longer or more explicit than Pip.'' of the book Great Expectations, by Charles Dickens.}
\end{figure}

The modeling process starts by removing punctuation and words that convey little semantic content (see the Supplementary Information (SI)-Sec.1), such as articles and prepositions. Then, the remaining words are transformed into their canonical form, i.e. nouns and verbs are converted into the singular and infinitive forms, respectively. This step is performed using the MXPOST part-of-speech tagger~\cite{ratna}, which assists the resolution of ambiguities. The transformation to the canonical form (lemmatization) is done to cluster words referring to the same concept into a single node of the network despite the differences in flexion. At last, adjacent words in the written text are connected in the network according to the natural reading order (the left word is the source node and the right word is the target node). The modeling is demonstrated in Table \ref{tab.4} for the pre-processing steps, while Fig. \ref{fig:network} illustrates the network obtained from a small extract of the book Great Expectations, by Charles Dickens.

\begin{table}
\centering
\caption{\label{tab.4} Illustration of the pre-processing (removal of stopwords and punctuation marks) and lemmatization of the extract ``My father's family name being Pirrip, and my Christian name Philip, my infant tongue could make of both names nothing longer or more explicit than Pip.'' obtained from the book Great Expectations, by Charles Dickens.}
\begin{tabular}{@{}ccc}
\br
\textbf{Original}&\textbf{Without stopwords}&\textbf{After lemmatization}\\
\br
My father's family name & father family name &  father family name \\
Pirrip, and my , & Pirrip  & Pirrip \\
Christian name Philip & Christian name Philip & Christian name Philip \\
my infant tongue & infant tongue & infant tongue \\
could make of both     &   could make both &   can make both \\
names nothing longer &  names longer & name long \\
or more explicit than Pip & more explicit Pip & more explicit Pip \\
\br
\end{tabular}
\end{table}

\subsection{Complex Networks Measurements} \label{medidas}

\begin{figure}
    \begin{center}
    \includegraphics[width=0.85\textwidth]{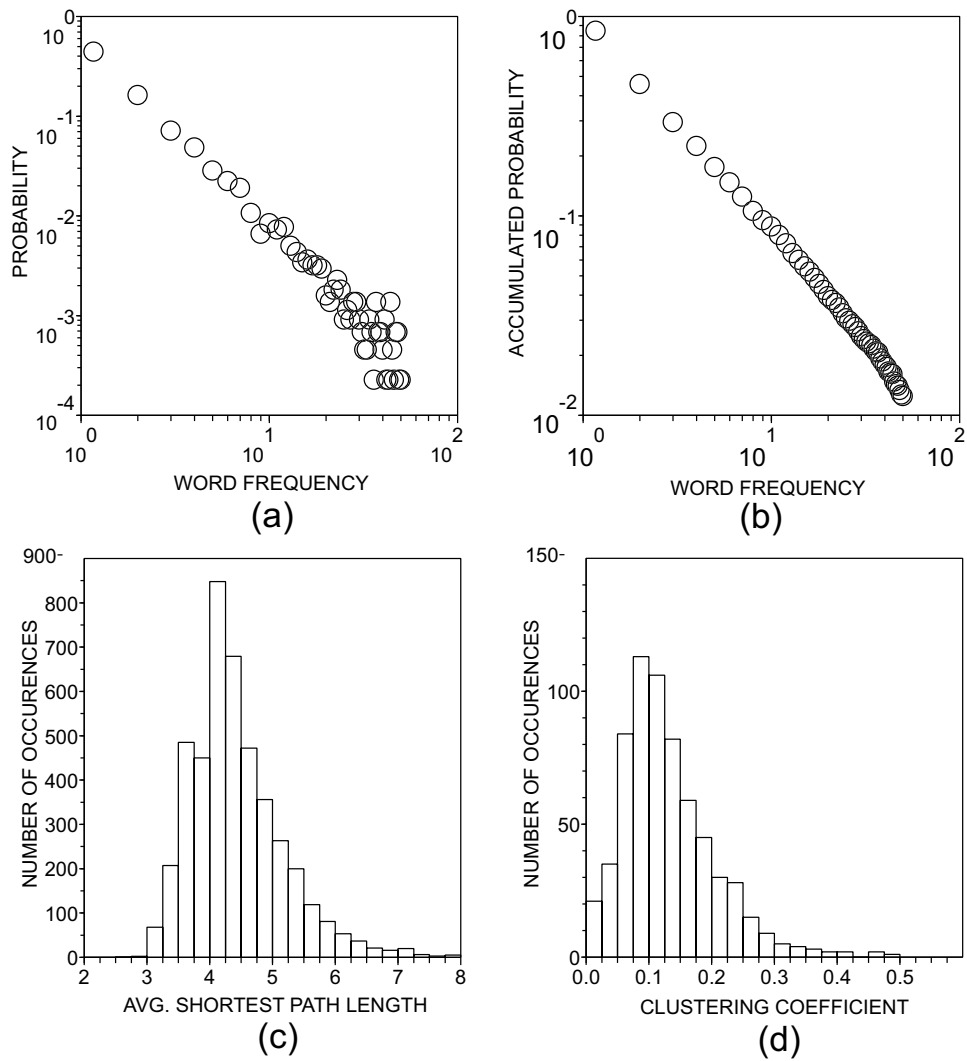}
    \end{center}
    \caption{\label{figDists}Example of distributions of measurements for the book Great Expectations, by Charles Dickens. The measurements used were: {(a) simple word frequency; (b) accumulated word frequency;} (c) average shortest path length; and (d) clustering coefficient. {The adjusted R-square found in (a) was equal to 0.9348, which confirms that the frequency distribution is very similar to a power law distribution.}}
    \end{figure}

Several metrics extracted from the networks were used to quantify the style of the books.
From each local measurement (i.e., which refers to a node) we derived some quantities describing the distribution of the networks in order to quantify the style of whole books. The measurements and their corresponding distribution descriptors were chosen because they have been useful to quantify the style of texts in previous studies~\cite{altmann}. The simplest measurement refers to the number $N$ of nodes in the network, which corresponds to the size of the vocabulary used to write the piece of text analyzed. The distribution of word frequency was characterized using the coefficient $\gamma$ of the frequency distribution $p_k$:
\begin{equation}
    p_k \sim c k^{-\gamma},
\end{equation}
where $c$ is a normalization constant (see Fig. \ref{figDists}(a) for an example of the frequency distribution $p_k$ of a specific book).
{We did not verify explicitly whether the degree obeys a power-law distribution because $k$ is proportional to the frequency of words. Since the word frequency follows the Zipf's Law~\cite{Manning,zipf}, the degree is guaranteed to obey a power-law distribution\footnote{{The power-law distribution was verified for all texts of the database}.}.
}
To compute $\gamma$, we employed a technique based on the accumulated distribution $p_k$ (see Fig. \ref{figDists}(b)) described in Ref.~\cite{bauke}. We also used the frequency of words (or equivalently the degree $k$ of the nodes) to calculate the assortativity $\Gamma$~\cite{assort1,assort2,eigen} (or degree-degree correlation) of the network as:
    \begin{equation}
    \Gamma = \frac{ \frac{1}{M}\sum_{j>i}^{}k_ik_ja_{ij} - \bigg{[} \frac{1}{M}\sum_{j>i}^{}\frac{1}{2}(k_i + k_j)a_{ij}\bigg{]}^2}{\frac{1}{M}\sum_{j>i}^{}\frac{1}{2}(k_i^2 + k_j^2)a_{ij} - \bigg{[} \frac{1}{M}\sum_{j>i}^{}\frac{1}{2}(k_i + k_j)a_{ij}\bigg{]}^2}
    \end{equation}
    where $M = 21,900$\footnote{To avoid effects from the size of the books, for obtaining the complex network we used only the first $M$ + 1 words of each book.} is the number of edges of the network and $a_{ij} = 1$ if nodes $i$ and $j$ are connected and $a_{ij} = 0$ otherwise. If positive values are obtained for $\Gamma$, then highly connected nodes are usually connected to other highly connected nodes, indicating that there may exist regions where nodes are highly interconnected~\cite{assort1}. Conversely, if $\Gamma$ is negative then highly connected nodes are commonly connected to little connected nodes.

    In addition to measurements based on the number of nodes of the network and on the degree, the distance between concepts was employed to characterize the structure of the books. This measurement, widely known in the theory of networks as average shortest path length $l$~\cite{newmanbook}, is calculated from the distance $d_{ij}$, which represents the minimum cost (minimum number of edges) required to reach node $j$, starting from node $i$. After computing all pairs of values $d_{ij}$, the average shortest path length $l_i$ of each node $i$ is:
    \begin{equation} \label{eq:sp}
	l_i = \frac{1}{N-1} \sum_{j \neq i}^{} d_{ij}.
    \end{equation}
    Since $l_i$ is defined for each node individually, the network is characterized by a distribution of $l_i$ (see the distribution of $l_i$ for a specific book in Fig. \ref{figDists}(c)). The distribution was characterized quantitatively by computing the average $\langle l \rangle$ and standard deviation $\Delta l$. Additionally, we computed the weighted average $(1 / \sum k_i) \sum k_i l_i \equiv \langle l_w \rangle$, so that greater importance was given to the most frequent words in the text. The third moment $\varsigma(l)$
    \begin{equation}
        \varsigma(l) = \frac{1}{N} \sum_{i=1}^{N} \Bigg{(} \frac{l_i - \overline{l}}{\Delta l}\Bigg{)}^3 = \frac{1}{N (\Delta l)^3} \Bigg{(} \sum_{i=1}^{N} l_i^3 - 3 \overline{l} \sum_{i=1}^{N} l_i^2 + 2 N \overline{l} ^ 3 \Bigg{)}
    \end{equation}
    was also computed.

    The last metric was the clustering coefficient ($C$)~\cite{newmanbook}, which quantifies the density of connections between the neighbors of a node $i$ according to:
    \begin{equation} \label{aglomeracao}
	C_i = \frac{ 3 \sum_{k > j > i} a_{ij} a_{ik} a_{jk} }{ \sum_{k > j > i} a_{ij} a_{ik} + a_{ji} a_{jk} + a_{ki} a_{kj} }.
    \end{equation}
    The clustering coefficient in equation \ref{aglomeracao} represents the fraction of the number of triangles among all possible connected sets of three nodes, and therefore $0 \leq C_i \leq 1$. Similarly to the average shortest path length, it is also necessary to quantitatively characterize the distribution of the measurement (see an example of distribution of $C$ in Fig. \ref{figDists}(d)). We therefore computed the average $\langle C \rangle$, the standard deviation $\Delta C $, the weighted average $(1 / \sum k_i) \sum k_i C_i \equiv \langle C_w \rangle$ and the third moment $\varsigma(C)$ to characterize the distribution. 

    \section{Database} \label{ssec.books}

    The database comprises $77$ books available online at the Gutenberg project repository~\cite{gutemberg}, whose publication date ranged from $1590$ to $1922$. Tables S1-S3 in (SI)-Sec.2 give the details of the books. The texts were represented with complex networks~\cite{cancho2,summ,trad1,trad2,cancho,sole,fatoficcao,patt,lantiq1,poesias,masucci}, in which the edges are defined on the basis of co-occurrence of words (see Sec.~\ref{modelamento}). The latter procedure has been proven suitable to quantify both the style and structure of texts (see e.g. Refs.~\cite{trad2,fatoficcao,poesias}). The details of the procedures adopted to model texts as complex networks and a description of the measurements employed to characterize the networks are given in Section \ref{modelamento}.

    \section{Results and Discussion}

\begin{figure}[!ht]
    \begin{center}
        \includegraphics[width=0.85\textwidth]{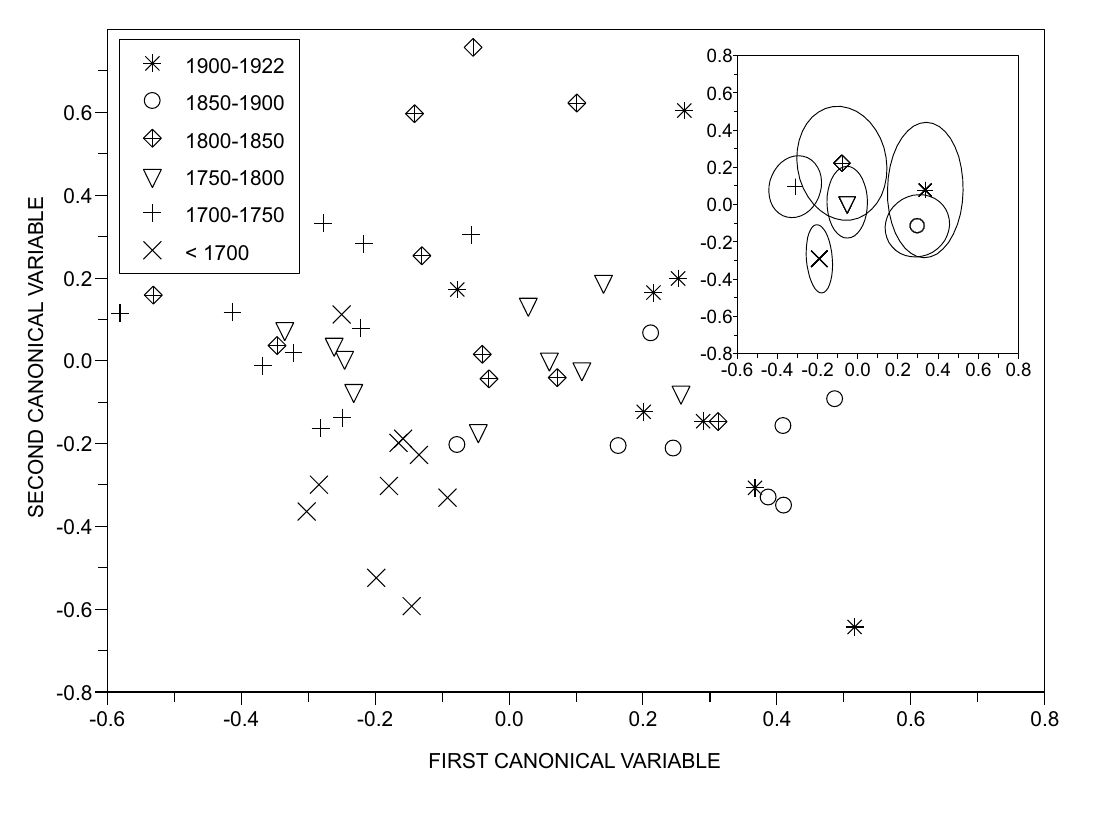}
    \end{center}
    \caption{\label{fig1.s1} { Scatter plot (CVA projection) representing the style of each book using $6$ literary styles}. Each style is represented by a set of $10$ books. The inset displays the dispersion of the literary styles.}
\end{figure}

    The evolution of literary styles was quantified considering the $11$ measurements from complex networks described in Sec.~\ref{medidas} for the books from the Project Gutenberg~\cite{gutemberg}. The main measurements were the shortest path length ($l$), the clustering coefficient ($C$), the assortativity ($\Gamma$), the power law coefficient of the degree distribution ($\gamma$) and the size of the vocabulary ($N$). An initial, arbitrary division of the books in $6$ intervals of $50$ years, according to their publication date, led to the clusters shown in the Canonical Variate Analysis (CVA, see details in (SI)-Sec.3) plot in Fig. \ref{fig1.s1}. The distinction was relatively poor, especially considering the standard variation ellipses~\cite{lee} in the inset of the figure. Good separation was only possible when distant periods in time were compared, as their ellipses did not overlap. This difficulty in distinguishing literary movements should perhaps be expected as there is no reason for sharp transitions to occur only because half century marks were reached. We also verified the distinguishability of clusters with the Principal Component Analysis (PCA, see (SI)-Sec.3), but the distinction was also poor.

\begin{figure}[!ht]
    \begin{center}
        \includegraphics[width=0.85\textwidth]{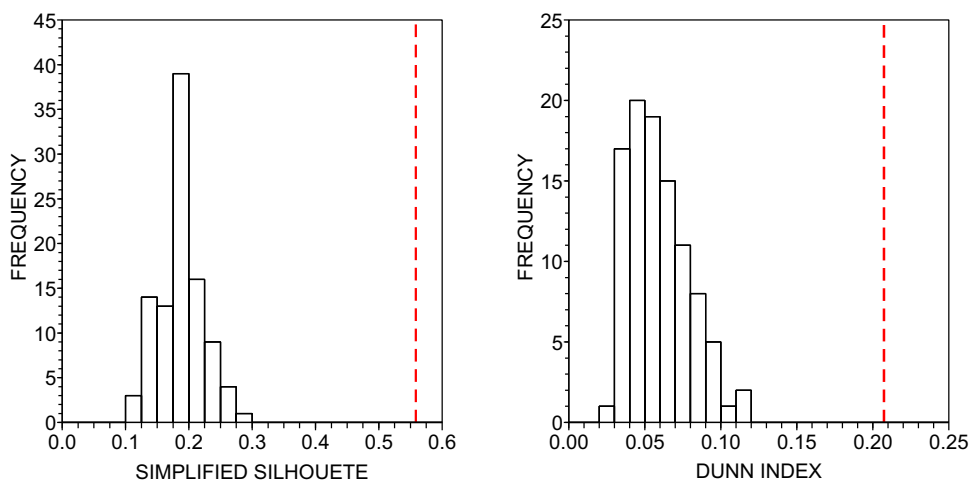}
    \end{center}
    \caption{\label{fig2sig} {Significance test performed for (a) the simplified silhouette and for (b) the Dunn Index. The histograms represent the values of the cluster quality indices considering a random distribution of points and the dotted lines represent the clustering quality indices obtained for the clustering illustrated in Figure \ref{fig2}. Because the silhouette for the random case $SWC_{rand} = 0.187 \pm 0.036$ is smaller than the silhouette $SWC = 0.558$ for the clustering of Figure \ref{fig2}, the clustering inferred is significant. The same applies for the Dunn index because $DN_{rand} = 0.059 < DN = 0.207$.}}
\end{figure}

\begin{figure}[!ht]
    \begin{center}
        \includegraphics[width=0.85\textwidth]{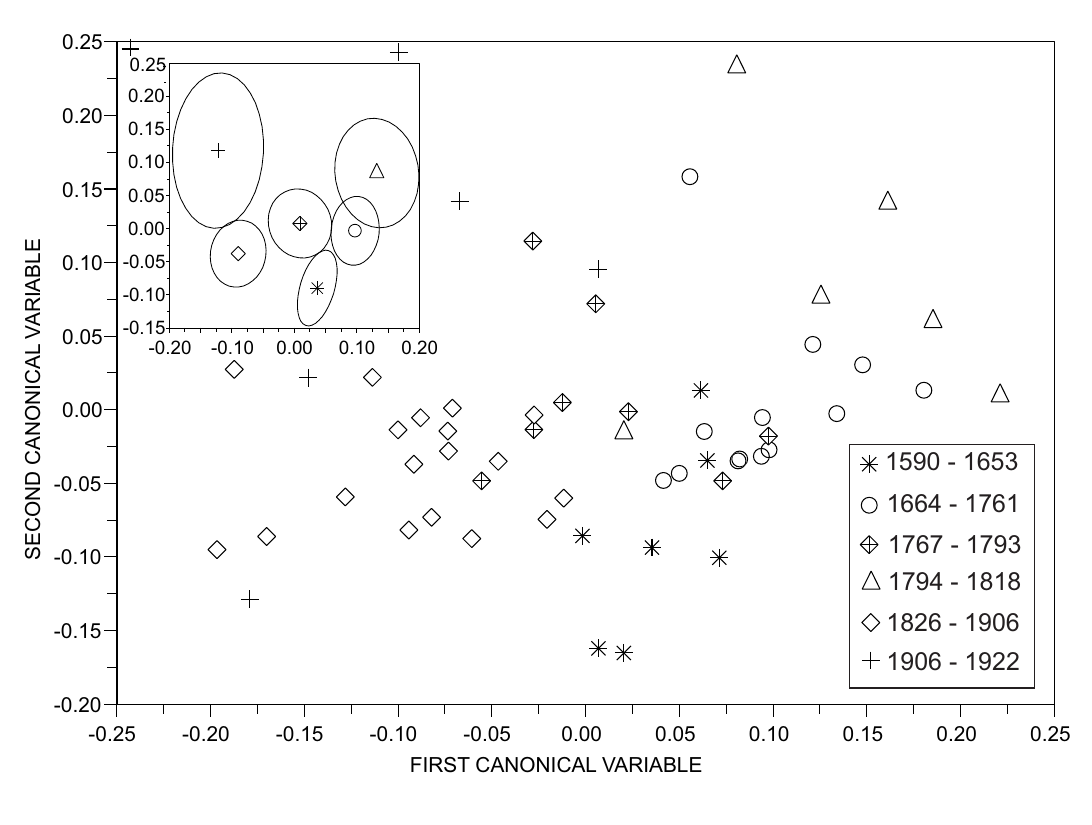}
    \end{center}
    \caption{\label{fig2} { Scatter plot representing the best clustering considering the writing style}. Note that besides being a good partitioning scheme, it also keeps a good representation of the original database, since $82$~\% of the variance are kept in the CVA projection.}
\end{figure}

In order to verify whether books from distinct publication dates could be distinguished at all, we adopted a systematic procedure for the partition of the dataset using an optimization approach. This was performed by assessing the quality of the clustering under the condition that books with consecutive publication dates should belong either to the same cluster or lie in the boundaries of consecutive clusters. More specifically, we varied the delimiters and number of clusters in the database and quantified the quality of the clustering using $2$ indices, viz. the simplified silhouette (SWC) and the Dunn index (DN) (see (SI)-Sec.4). Good distinction of writing styles was obtained for {$3$, $4$, $5$, $6$ and $7$ clusters (see Figure S1 of the SI)}, according to the two indices (SWC and DN). {The best partition, which was found to be statistically significant (see Figure \ref{fig2sig})}, was obtained with SWC and CVA projection, leading to the $6$ clusters in Fig. \ref{fig2}, where there is almost no overlap among clusters, as shown in the inset. Most significantly, the $6$ time periods inferred from this analysis coincide with well-established literary movements listed in Table \ref{tab.literario}.

\begin{table}[!ht]
\centering
\caption{Relationship between the best clustering of writing styles the traditional classification of literary movements.}
\begin{tabular}{@{}cccc}
  \br
  \textbf{Cluster Boundary}      & \textbf{Literary Boundary} & \textbf{Literary Movement}    & \textbf{Reference}      \\
  \br
  1590 - 1653   &   1558 - 1603  &  Elizabethan era                 & ~\cite{wiki1}                \\
  1664 - 1761   &   1660 - 1798  &  Neoclassicism/Enlightenment     & ~\cite{wiki2,wiki3,wiki4}    \\
  1767 - 1793   &   1660 - 1798  &  Neoclassicism/Enlightenment     & ~\cite{wiki2,wiki5}          \\
  1794 - 1818   &   1764 - 1820  &  Gothic fiction                  & ~\cite{wiki2,wiki5}          \\
  1826 - 1906   &   1830 - 1900  &  Realism                         & ~\cite{wiki2}                \\
  1826 - 1906   &   1865 - 1900  &  Naturalism                      & ~\cite{wiki2,wiki6}          \\
  1906 - 1922   &   1890 - 1940  &  Modernism                       & ~\cite{wiki2,wiki7}          \\
  \br
\end{tabular}
\label{tab.literario}
\end{table}

\begin{figure}[!ht]
    \begin{center}
        \includegraphics[width=0.7\textwidth]{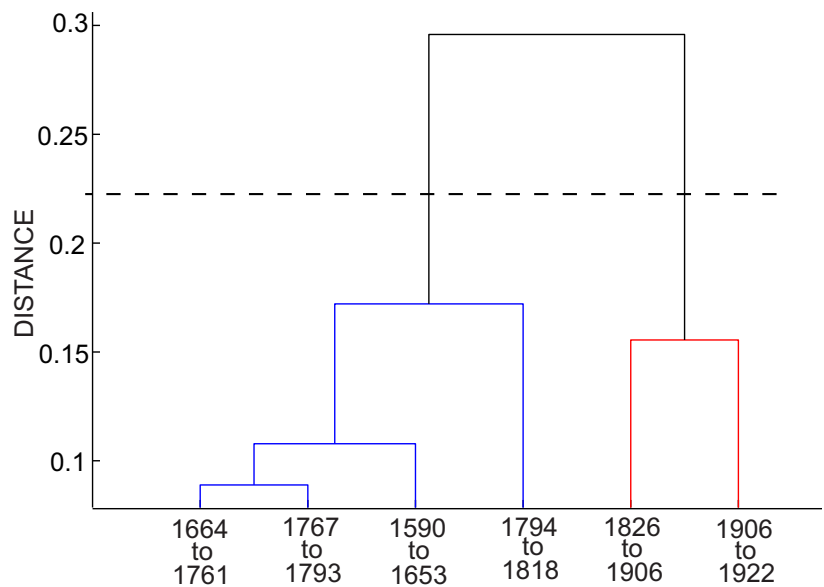}
    \end{center}
    \caption{\label{dendo2} { Hierarchical relationship between literary periods using the Wards linkage strategy}. The $2$ groups after the division performed with a particular threshold (dotted line) corresponds to the oldest and to the newest books. }
\end{figure}

Other important features are inferred from Fig. \ref{fig2}. First, clusters for subsequent time periods are normally placed next to each other, indicating smooth changes in writing style over time. The same conclusion can be inferred from the analysis of the hierarchical clustering in Fig. \ref{dendo2} with the Wards~\cite{ward} distance. The exception to this trend was the major change from the $1794 - 1818 \rightarrow 1826 - 1906$ period, which may be the consequence of a drastic change in style triggered by the French Revolution ($1789$). As for the variance among clusters, the lowest and highest values applied to the $1590 - 1653$ and $1906 - 1922$ periods, respectively. These results are intuitive as little change in style could be expected in older periods, while in the recent periods less uniformity could be the result of the coexistence of many writing styles.

\begin{figure}[!ht]
       \begin{center}
           \includegraphics[width=1\textwidth]{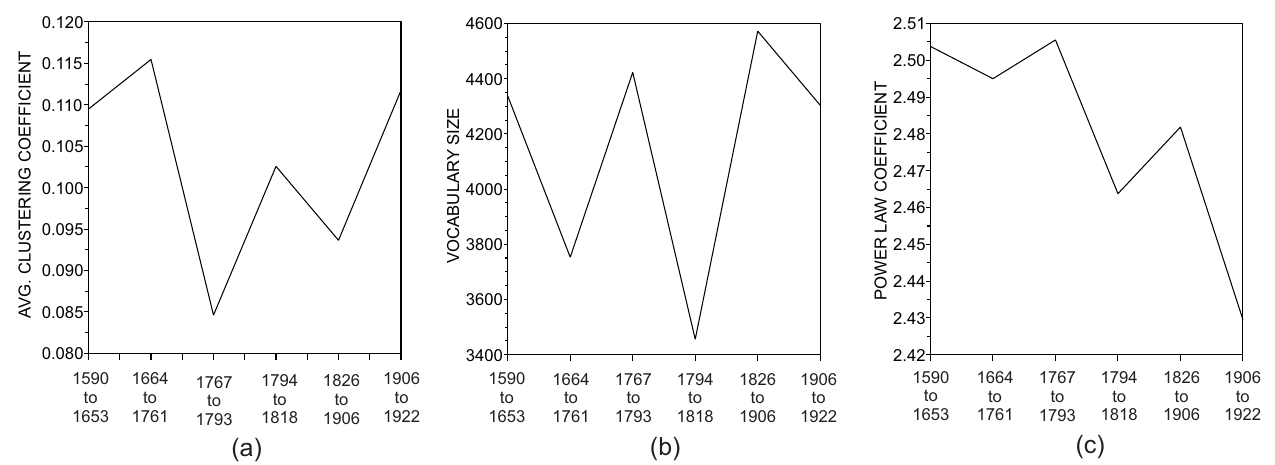}
       \end{center}
       \caption{\label{fig3} { Dynamics of (a) average clustering coefficient; (b) vocabulary size; and (c) coefficient of the power law}. While the clustering coefficient and the vocabulary size oscillate throughout the periods, the coefficient of the power law tends to decrease, which shows that words were used in a more uniform way in the later periods.}
\end{figure}

{
The most important factors contributing to the separation of literary styles were determined in two distinct ways. The first technique considered a feature to be relevant if it was capable of providing significant distinction between groups, regardless of the other features. The list of metrics and the corresponding p-value for the difference of a given measurement between pairs of clusters are given in Table \ref{tab.relevance}. The asymmetry in the distribution of the average shortest path length $\varsigma(l)$ and the vocabulary size $N$ exhibited the most significant variations. Interestingly, similar results were reported in Ref.~\cite{altmann}, where these two measurements were also useful to characterize personal writing styles. In the second evaluation, a feature was considered relevant if it was able to provide good distinction between groups based on the interdependencies of features. This evaluation was carried out by
}
computing the importance of each measurement for the axes in the CVA plots. The results in Tables \ref{tab.s5} and \ref{tab.ss5} point to the clustering coefficient ($C$ and $C_w$) as the main factor for the distinction in $6$ clusters. Since there is evidence that the clustering coefficient quantifies whether words are restricted to specific or generic contexts (an explanation of this property is given in Ref.~\cite{altmann})\footnote{Context-specific restricted words are those appearing in only a few contexts. For example, the concept ``teacher'' usually induces concepts related to the learning environment. On the other hand, generic words may appear in a myriad of situations. Examples are ``red'' (red car, red wall or red skin) and ``identical'' (identical behaviors, identical grades or identical plates}, it seems that the extent of use of generic or specific words varied along history. This change has not been monotonic, as indicated in Fig. \ref{fig3}(a). In fact, most of the network measurements fluctuated over time, including the size of the vocabulary, whose considerable change was responsible for the most drastic transition, from the $1794 - 1818 \rightarrow 1826 - 1906$ periods. This is clearly illustrated in Fig. \ref{fig3}(b). The only metric with a well-defined trend over time was the coefficient of the power law for the scale-free networks representing the texts. The decreasing trend in Fig. \ref{fig3}(c) points to a smoother, and therefore more uniform, frequency distribution, which means that the difference in frequency between low and high-frequency words decreased with time.

\begin{table}[!ht]
\centering
\caption{{List of the most significant transitions. Taken individually, the most prominent measurements for discriminating between clusters are the size of the vocabulary $N$ and the third moment of the average shortest path length $\varsigma(L)$}.}
\begin{tabular}{@{}cccc}
  \br
  \textbf{Measurement} & {\bf Feature}  & \textbf{Transition}     &  {\bf p-value} \\
  \br
  Vocabulary & $N$ &    $1590 - 1653$ $\rightarrow$ $1794 - 1818$ & 0.048 \\
             & $N$ &    $1664 - 1761$ $\rightarrow$ $1767 - 1793$ & 0.051 \\
             & $N$ &    $1664 - 1761$ $\rightarrow$ $1826 - 1906$ & 0.001 \\
             & $N$ &    $1767 - 1793$ $\rightarrow$ $1794 - 1818$ & 0.011 \\
             & $N$ &    $1794 - 1818$ $\rightarrow$ $1826 - 1906$ & $< 1.0~10^{-3}$ \\
  \br
  Assortativity & $\Gamma$ & $1590 - 1653$ $\rightarrow$ $1767 - 1793$ & 0.008 \\
                & $\Gamma$ & $1590 - 1653$ $\rightarrow$ $1826 - 1906$ & 0.044 \\
                & $\Gamma$ & $1664 - 1761$ $\rightarrow$ $1767 - 1793$ & 0.041 \\
                & $\Gamma$ & $1664 - 1761$ $\rightarrow$ $1826 - 1906$ & 0.006 \\
  \br
  Shortest Path & $\langle l \rangle$        & $1664 - 1761$ $\rightarrow$ $1826 - 1906$ & 0.049 \\
                & $\langle l_w \rangle$      & $1664 - 1761$ $\rightarrow$ $1906 - 1922$ & 0.050 \\
                & $\Delta L$ & $1590 - 1653$ $\rightarrow$ $1906 - 1922$ & 0.031 \\
                & $\Delta L$ & $1664 - 1761$ $\rightarrow$ $1906 - 1922$ & 0.022 \\
                & $\Delta L$ & $1767 - 1793$ $\rightarrow$ $1906 - 1922$ & 0.023 \\
                & $\Delta L$ & $1826 - 1906$ $\rightarrow$ $1906 - 1922$ & $< 1.0~10^{-3}$ \\
                & $\varsigma(l)$ & $1590 - 1653$ $\rightarrow$ $1826 - 1906$ & 0.028 \\
                & $\varsigma(l)$ & $1590 - 1653$ $\rightarrow$ $1906 - 1922$ & $< 1.0~10^{-3}$ \\
                & $\varsigma(l)$ & $1664 - 1761$ $\rightarrow$ $1906 - 1922$ & $< 1.0~10^{-3}$ \\
                & $\varsigma(l)$ & $1767 - 1793$ $\rightarrow$ $1906 - 1922$ & $0.001$ \\
                & $\varsigma(l)$ & $1794 - 1818$ $\rightarrow$ $1906 - 1922$ & $0.019$ \\
                & $\varsigma(l)$ & $1826 - 1906$ $\rightarrow$ $1906 - 1922$ & $< 1.0~10^{-3}$ \\
  \br
  Clustering    & $\langle C \rangle$ & $1664 - 1761$ $\rightarrow$ $1767 - 1793$ & 0.048 \\
                & $\langle C \rangle$ & $1664 - 1761$ $\rightarrow$ $1826 - 1906$ & 0.051 \\
                & $\langle C_w \rangle$ & $1664 - 1761$ $\rightarrow$ $1767 - 1793$ & 0.054 \\
                & $\langle C_w \rangle$ & $1664 - 1761$ $\rightarrow$ $1826 - 1906$ & 0.055 \\
                & $\Delta C$ & $1664 - 1761$ $\rightarrow$ $1767 - 1793$ & 0.054 \\
                & $\varsigma(C)$ & $1590 - 1653$ $\rightarrow$ $1767 - 1793$ & $0.045$ \\
  \br
\end{tabular}
\label{tab.relevance}
\end{table}

\begin{table}[!ht]
\centering
\caption{Importance of each measurement for the first canonical variable, where the clustering coefficient $C$ and the average shortest path length $l$ were the most prominent.}
\begin{tabular}{@{}cc}
  \br
  \textbf{Measurement}      & \textbf{Prominence} \\
  \textbf{(First Axis)}     & \textbf{(First Axis)}  \\
  \br
    $\langle C_w \rangle$  &   33.3~\%  \\
    $\langle C \rangle$    &   31.6~\%  \\
    $\Delta C$             &    6.6~\%  \\
    $\langle l \rangle$    &    6.4~\%  \\
    $\Gamma$               &    5.1~\%  \\
  \br
\end{tabular}
\label{tab.s5}
\end{table}

\begin{table}[!ht]
\centering
\caption{Importance of each measurement for the second canonical variable, where the clustering coefficient $C$ and the average shortest path length $l$ were the most prominent.}
\begin{tabular}{@{}cc}
  \br
  \textbf{Measurement}    & \textbf{Prominence}      \\
  \textbf{(Second Axis)}  & \textbf{(Second Axis)}   \\
  \br
    $\langle C \rangle$           & 34.5~\%   \\
    $\langle C_w \rangle$         & 33.7~\%   \\
    $\langle l_w \rangle$         &  9.5~\%   \\
    $\langle l \rangle$           &  9.4~\%   \\
    $\Delta C$                    &  3.4~\%   \\
  \br
\end{tabular}
\label{tab.ss5}
\end{table}

The changes in style between any two consecutive clusters appeared to have been driven by opposition~\cite{philo} {(see Appendix A)}, which quantifies the extent into which the current period can be thought of as an opposite movement to the previous literary movements.
The coefficient satisfies the inequality $W_{ij} > 0$, with the exception of the $1826 - 1906 \rightarrow 1909 - 1922$ transition. Furthermore, the opposition movement was more significant than the skewness movement $s_{ij}$ {(see Appendix A)}, which quantifies how much the change in the current style deviates from the opposition movement.
The results are given in Table \ref{tab.res1}. In other words, the innovation of style ($\overrightarrow{v_i}$, {see definition in Appendix A}) was generally driven by contrasting the previous styles ($\overrightarrow{a_i}$, {see definition in Appendix A}). As for the dialectics $\rho_{ijk}$ {(see Appendix A)}, which quantifies how the current movement $i$ is an implication of the two previous movements $j$ and $k$, no clear pattern could be identified in Table \ref{tab.res2}. The lowest $\rho_{ijk}$ (and therefore with the highest dialectics) appeared during the $19$th century. Thus, realism is a literary style that better approximates as a synthesis of the two previous literary periods.

\begin{table}[!ht]
\centering
\caption{Opposition ($W_{ij}$) and skewness ($s$) indices.}
\begin{tabular}{@{}ccc}
  \br
  \textbf{Period} & \textbf{$W_{ij}$} & \textbf{$s_{ij}$} \\
  \br
  1590 - 1653 $\rightarrow$ 1664 - 1761 & 1.00  &  0.00 \\
  1664 - 1761 $\rightarrow$ 1767 - 1793 & 0.39  &  0.08 \\
  1767 - 1793 $\rightarrow$ 1794 - 1818 & 0.35  &  0.18 \\
  1794 - 1818 $\rightarrow$ 1826 - 1906 & 1.09  &  0.07 \\
  1826 - 1906 $\rightarrow$ 1909 - 1922 & -0.01 &  0.08 \\
  \br
\end{tabular}
\begin{flushleft}
\end{flushleft}
\label{tab.res1}
\end{table}

\begin{table}[!ht]
\centering
\caption{Counter Dialectics index $\rho_{ik}$.}
\begin{tabular}{@{}cc}
  \br
  \textbf{Period}   &   $\rho_{ik}$   \\
  \br
  $1590 - 1653  \rightarrow 1664 - 1761 \rightarrow 1767 - 1793$  & 0.76 \\
  $1664 - 1761  \rightarrow 1767 - 1793 \rightarrow 1794 - 1818$  & 1.49 \\
  $1767 - 1793  \rightarrow 1794 - 1818 \rightarrow 1826 - 1906$  & 0.39 \\
  $1794 - 1818  \rightarrow 1826 - 1906 \rightarrow 1909 - 1922$  & 0.69 \\
  \br
\end{tabular}
\begin{flushleft}
\end{flushleft}
\label{tab.res2}
\end{table}

\begin{figure}[!ht]
\begin{center}
\includegraphics[width=0.5\textwidth]{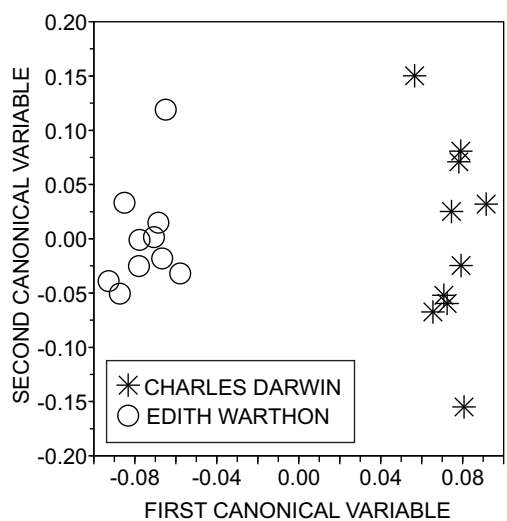}
\end{center}
\caption{\label{fig4} { Comparing Darwin's and Edith Warthon's styles with CVA projection}. A good separation can be observed indicating that these two authors had quite different styles.}
\end{figure}

In subsidiary studies we verified that the complex network metrics used are indeed efficient in distinguishing styles. For that we examined the writing style dynamics of $10$ books\footnote{The list of books is shown in Table S3 in (SI)-Sec.2.} of Charles R. Darwin ($1809$-$1882$) and Edith Wharton ($1862$-$1937$), whose styles are known to differ considerably. Indeed, this is confirmed in the CVA plot in Fig. \ref{fig4}, where again the most contributing factor for distinction was the clustering coefficient $C$, since both $\langle C \rangle$ and $\langle C_w \rangle$ are responsible for 44 \% of the weights in the first canonical variable axis.

\section{Conclusion and further work}

Changes in the writing style could be studied objectively by analyzing the metrics from complex networks representing texts from books published over several centuries. Significantly, the most appropriate clustering of books matched the traditional literary classification, with the most contributing factor for distinguishability being the {average shortest path length}. {We found it to be possible to distinguish literary movements using only the vocabulary size or the asymmetry of the average shortest path length distribution. Innovation in writing style was found to be driven mainly by opposition, with growing trend of literary development toward counter-dialectics. Interestingly, these findings represent the generalization of previous results where a dependence was established between network topology and style of machine translations~\cite{trad1,trad2} and style of authors~\cite{altmann}. We believe that the approach used here may be useful to study the evolution of any system of interest, since the basic concepts (i.e. characterization through features and use of time series) are completely generic.}

{As future work, we plan to employ additional complex network measurements in a larger database to verify if the discrimination can be further improved. We shall also examine the relationship between semantics and topology, by generating clusters using the semantics of words to be compared with the clusters obtained from the analysis of network topology. A more challenging endeavor will be to extend the study to other languages, in order to probe whether the patterns revealed in this paper can be generalized.}

\section*{Acknowledgments}
The authors are grateful to FAPESP (2010/00927-9) and CNPq (Brazil) for the financial support.

\newpage

\section*{Appendix A - Mathematical quantification of writing style}

\begin{figure}[!ht]
    \begin{center}
    \includegraphics[width=0.8\textwidth]{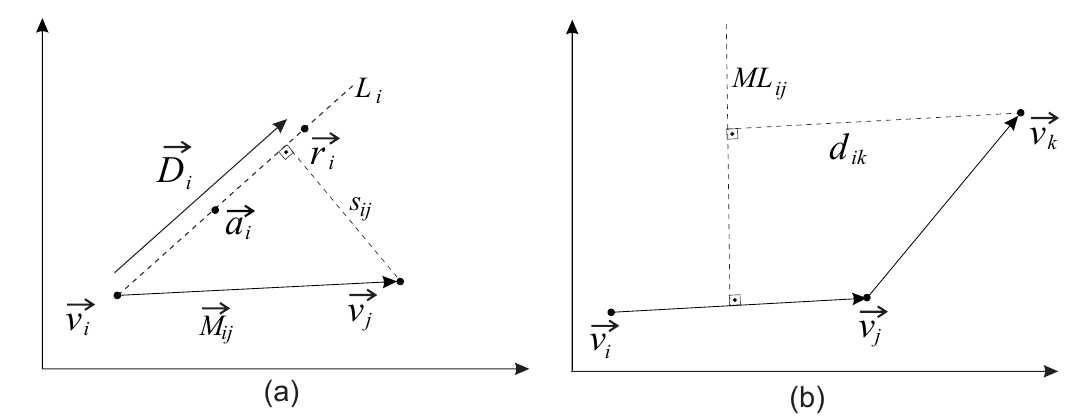}
    \end{center}
    \caption{\label{ilu1} {Illustration of the quantities employed to define the {\it opposition}, {\it skewness} and {\it counter-dialectics} indices.}}
    \end{figure}

In this appendix we quantify mathematically  the variation of writing style.
To quantify the change in style over time, we used three concepts, namely \textit{opposition index}, \textit{skewness index} and \textit{counter-dialectics index}, which depend on the measurements computed in each step of the temporal series. {For each element $i$ of the temporal series, which represents the value for the measurements described in Sec.~\ref{medidas}, we defined the $11$-dimensional vector $\overrightarrow{v_i}$:}
{
\begin{equation}
    \overrightarrow{v_i} = \Big{[}~N~\Gamma~\gamma~\langle C \rangle~\langle C_w \rangle~\Delta C~\varsigma(C)~\langle l \rangle~\langle l_w \rangle~\Delta l~ \varsigma(l)~\Big{]}^T.
\end{equation}}
The large amount of data generated were visualized by projecting $\overrightarrow{v_i}$ into a two dimensional space before computing the indices, and this also helped to remove undesirable correlations. The projection techniques employed are described in (SI)-Sec.3. Using the projected $\overrightarrow{v_i}$, and considering $t$ elements in the time series, $\overrightarrow{a_i}$ was defined the average state at time $i$, $i \leq t$ as:
    \begin{equation}
        \overrightarrow{a_i} = \frac{1}{i} \sum_{j=1}^{i} \overrightarrow{v_i}.
    \end{equation}
    Given $\overrightarrow{a_i}$, the \textit{opposite state} of the current state $i$ (see Fig. \ref{ilu1}(a)) for a geometrical interpretation) is given by:
    \begin{equation}
        \overrightarrow{r_i} = \overrightarrow{v_i} + 2(\overrightarrow{a_i} - \overrightarrow{v_i}) = 2\overrightarrow{a_i} - \overrightarrow{v_i},
    \end{equation}
    and given $\overrightarrow{r_i}$ and $\overrightarrow{v_i}$, the \textit{opposition vector} $\overrightarrow{D_i}$ of state $\overrightarrow{v_i}$ (see Fig. \ref{ilu1}(a) is given by:
     \begin{equation}
        \overrightarrow{D_i} = \overrightarrow{r_i} - \overrightarrow{v_i}.
    \end{equation}
    For two consecutive books $i$ and $j$, the vector representing the style change $\overrightarrow{M_{ij}}$ (see Fig. \ref{ilu1}(a)) is:
    \begin{equation}
        \overrightarrow{M_{ij}} = \overrightarrow{r_i} - \overrightarrow{v_i}.
    \end{equation}
    The vector $\overrightarrow{M_{ij}}$ is important because its norm $\| \overrightarrow{M_{ij}} \|$ quantifies the change in style in relation to the previous state $\overrightarrow{v_i}$. With $\overrightarrow{M_{ij}}$, the \textit{opposition index} $W_{ij}$ is the component of $\overrightarrow{M_{ij}}$ over $\overrightarrow{D_i}$:
    \begin{equation}
        W_{ij} = \frac{\overrightarrow{M_{ij}} \cdot \overrightarrow{D_i} }{\|\overrightarrow{D_i}\|^2}
    \end{equation}

    If the current style tends to oppose the previous one, then the component of $\overrightarrow{M_{ij}}$ over $\overrightarrow{D_i}$  will have a high value. This quantifier is useful, for example, to identify little stylistic innovation: if opposite movements are repeated over and over again, then there is no innovation at all.

    The {\it skewness index} $s_{ij}$, which is depicted in Fig. \ref{ilu1}(a), is defined as the distance between $\overrightarrow{v_j}$ and the line defined by $\overrightarrow{D_i}$. This index quantifies how far the stylistic movement is from the opposite movement. It is useful to identify trivial oscillations within the line $L_i$, for in this case a series of movements with zero \textit{skewness index} would be observed.

    The dialectics between three consecutive styles $i$, $j=i+1$ and $k=j+1=i+2$ in the temporal series was quantified as follows. {If $\overrightarrow{v_k}$ is the outcome of a synthesis of the styles represented by $\overrightarrow{v_i}$ and $\overrightarrow{v_j}$, then the distance $d_{ik}$ between $\overrightarrow{v_k}$ and the middle line $ML_{ij}$ defined by $\overrightarrow{v_i}$ and $\overrightarrow{v_j}$ (see Fig. \ref{ilu1}(a)) will be small. The {\it counter dialectics index}\footnote{Note that we referred to $\rho_{ik}$ as {\it counter dialectics index} instead of {\it dialectics index} because it is defined as a distance. Hence, there is an inverse proportion between $\rho_{ik}$ and the concept of dialectics.} $\rho_{ik}$ is}:
    \begin{equation}
        \rho_{ik} = \frac{d_{ik}}{\|M_{ij}\|}
    \end{equation}

    Further details regarding the definition of the opposition $W_{ij}$, sknewness $s_{ij}$ and counter-dialetics $\rho_{ik}$ are given in Ref.~\cite{philo}.

\end{document}